\documentclass[12pt,preprint]{aastex}
\usepackage{natbib}
\shorttitle{Jitter Radiation and GRB Polarization}

\shortauthors{Mao et al.}

\begin{document}


\title{Application of Jitter Radiation: Gamma-ray Burst Prompt Polarization}


\author{
Jirong Mao\altaffilmark{1,2,3} and
Jiancheng Wang\altaffilmark{2,3}
}
\altaffiltext{1}{Astrophysical Big Bang Lab, RIKEN, Saitama 351-0198, Japan}
\altaffiltext{2}{Yunnan Observatory, Chinese Academy of Sciences, 650011 Kunming, Yunnan Province, China}
\altaffiltext{3}{Key Laboratory for the Structure and Evolution of Celestial
Objects, Chinese Academy of Sciences, 650011 Kunming, China}

\email{jirong.mao@riken.jp}

\begin{abstract}
A high-degree of polarization of gamma-ray burst (GRB) prompt emission has been confirmed in recent years. In this paper,
we apply jitter radiation to study the polarization feature of GRB prompt emission. In our framework, relativistic electrons
are accelerated by turbulent acceleration. Random and small-scale magnetic fields are generated by turbulence.
We further determine that the polarization property of GRB prompt emission is governed by the configuration of the random and 
small-scale magnetic fields. A two-dimensional compressed slab, which contains stochastic magnetic fields, is applied in our 
model. If the jitter condition is satisfied, the electron deflection angle in the magnetic field is very small and the electron trajectory can be treated as a straight line.
A high-degree of polarization can be achieved when the angle between the line of sight and the slab plane is small.
Moreover, micro-emitters with mini-jet structure are considered to be within a bulk GRB jet. The jet ``off-axis'' effect is
intensely sensitive to the observed polarization degree.
We discuss the depolarization effect on GRB prompt emission and afterglow. We also speculate
that the rapid variability of GRB prompt polarization
may be correlated with the stochastic variability of the turbulent dynamo or the magnetic reconnection of plasmas.
\end{abstract}




\section{Introduction}
One of the important properties of celestial radiation is polarization.
Polarization is produced by relativistic electrons emitting in magnetic fields and
can be detected by either high-energy satellites or ground-based telescopes. Through
this kind of polarization research, we can investigate both the radiation mechanisms and the magnetic field
characteristics of celestial objects.

Gamma-ray bursts (GRBs) are the most energetic explosions in the universe. Some polarization
detections of GRBs in the prompt $\gamma$-ray band were performed. A linear polarization with a degree of
$\Pi=80\%\pm 20\%$ in GRB 021206 was detected by {\it RHESSI} \citep{coburn03}. GRB 041219A,
observed by the International Gamma-Ray Astrophysics Laboratory, also has a high degree of
polarization. Values of $\Pi=98\% \pm 33\%$ and $\Pi=63\%\pm 30\%$ were reported by \citet{kalemci07}
and \citet{mcglynn07}, respectively. Recently, $\gamma$-ray prompt polarizations of three GRBs were detected by the
GRB polarimeter onboard {\it IKAROS}: GRB 100826A has an average polarization
degree of $27\% \pm 11\%$ \citep{yonetoku11}; GRB 110301A and GRB 110721A have high polarization
degrees of $70\% \pm 22\%$ and $84^{+16}_{-28}\%$, respectively \citep{yonetoku12}. Meanwhile, theoretical models
are strongly required to constrain the physical origin of these highly polarized GRB prompt photons and to explore
possible magnetic field configurations.

Synchrotron radiation is one kind of emission from relativistic electrons in an ordered and large-scale magnetic field. In general, the linear polarization degree is given as $\Pi=(3\nu_S+3)/(3\nu_S+5)$, where $\nu_S$ is the synchrotron spectral index \citep{rybicki79}. Some comprehensive models of GRB polarization have been proposed. \citet{granot03} studied the polarization of prompt emission in GRB 021026 considering a jet structure. A polarization degree larger than $50\%$ can be produced in the case of an ordered magnetic field. A jet structure was also studied by \citet{ghisellini99} and \citet{waxman03}. About a few percent of the polarization in GRB optical afterglows can be produced when a tangled magnetic field is introduced. \citet{toma09} presented a statistical study in X-ray band ($60-500$ keV) from Monte Carlo simulations. If the polarization detection ratio is larger than $30\%$ and $0.2<\Pi<0.7$, synchrotron radiation in an ordered magnetic field is the favored mechanism. If the polarization detection ratio is less than $15\%$, a random magnetic field may be possible.

If the magnetic field is small-scale and randomly oriented, the isotropic emission of relativistic electrons in such
a magnetic field cannot show any significant net polarization pattern. In this case, \citet{medvedev99}
propose a polarization origin from interstellar scintillation. If the emission was in some magnetic patches,
the measured polarization was estimated as $\Pi=\Pi_0/\sqrt{N}$, where $\Pi_0$ is the intrinsic polarization degree and $N$ is the patch number \citep{gruzinov99}. These studies indicate that low-degree polarization is related to tangled magnetic fields.
However, we note that \citet{lazzati09} obtained a highly polarized GRB from a fragmented fireball. The
pulse with high polarization, originating from a single fragment, is 10 times fainter than the brightest pulse of a GRB.

Despite some popular literature results on synchrotron polarization, we propose in this paper
an alternative possibility to explain the high degree of polarization of GRB prompt emission.
In general, jitter radiation originates from relativistic electrons accelerating in stochastic magnetic fields.
As these stochastic magnetic fields are randomly distributed, the electrons feel almost the same Lorentz force on average from different directions. Therefore, in the electron radiative plane, jitter radiation is highly symmetric and the polarization degree is nearly zero.
However, if the symmetric feature of radiation breaks down due to certain reasons, a strong polarization will appear in the jitter radiation.
In this paper, we apply a special magnetic field slab for the jitter radiation. Although the magnetic fields in this two-dimensional slab are randomly distributed, this slab provides an asymmetric configuration such that the jitter radiation is anisotropic in the radiative plane. Thus, the jitter radiation can be highly polarized.
There are two special physical points to our model: (1) a two-dimensional compressed slab as a particular magnetic field configuration, and (2) a bulk relativistic jet with a Lorentz factor $\Gamma_j$ and many micro-emitters with relativistic turbulent Lorentz factors $\Gamma_t$ within the bulk jet. Therefore, we expect the possibility of highly polarized GRB prompt emission due to the strongly anisotropic properties of the jitter radiation within a jet-in-jet structure.

In our scenario, with propagation and collision of internal shocks from the GRB central engine, turbulence is produced behind
the shock front. Random and small-scale magnetic fields can be generated by turbulence. Those electrons can be accelerated 
not only by diffusive shock acceleration but also by turbulent acceleration. The radiation mechanism of relativistic electrons radiating $\gamma$-ray photons in random and small-scale magnetic fields is the so-called
jitter radiation. As turbulence is important for particle acceleration, the electron energy distribution combines a power-law
shape and a Maxwellian shape. Due to the domination of random and small-scale magnetic fields, jitter photons come from those
micro-emitters with a jet structure. These mini-jets with turbulent Lorentz factors $\Gamma_t$ are within the bulk jet
with a Lorentz factor $\Gamma_j$. The total observed emission is all of the contributions from
these mini-jets. The framework of this scenario was previously built: a small-scale turbulent dynamo was realized by hydrodynamical simulations
\citep{sch04}, jitter radiation was presented \citep{medvedev00,medvedev06,kelner13}
and examined numerically \citep{sirony09,fre10}, the radiative synthetic spectra from relativistic shocks
were also simulated by \citet{martins09} and \citet{nishikawa11}, the radiation process in a sub-Larmor scale magnetic field
was re-examined by \citet{medvedev11}, the electron energy distribution was given by \citet{stawarz08} and \citet{giannios09},
turbulence-induced random and small-scale magnetic fields and related jitter radiation for GRB prompt emission were
explored by \citet{mao11}, this radiation spectrum is fully consistent with the high-frequency spectrum derived from numerical
calculations \citep{teraki11}, and GRB mini-jets were discussed by \citet{mao12}. Very recently, some detailed calculations of the microturbulent
dynamics behind relativistic shock fronts
\citep{lemoine13} and comprehensive analyses of pulses seen in {\it Swift}-Burst Alert Telescope GRB lightcurves \citep{bhatt12}
also shed light on the physics of jitter radiation, tangled magnetic fields, and relativistic turbulence.

We further apply our previous model \citep{mao11,mao12} to investigate GRB prompt
polarization in this paper. Magnetic field topology is essential for
jitter radiation properties \citep{reynolds10,reynolds12}. The configuration effect of random magnetic fields was presented by \citet{laing80}. In that work, a tangled magnetic field in a three-dimensional cube was compressed into a
two-dimensional slab. In the slab, the magnetic field is still random. The line of sight from an observer has a certain angle to the plane
of magnetic slab. Thus, the symmetric feature of random magnetic fields is broken. The compressibility properties of magnetic fields were studied in detail by \citet{hughes85}. The polarization from oblique
and conical shocks was given by \citet{cawthorne90} and \citet{nalewajko09}. Some numerical simulations of tangled
magnetic fields were also performed \citep{matthews90a,matthews90}. \citet{laing02} developed a calculation of
chaotic magnetic field compression. The polarization degrees and angles were put straightforward in some
cases. In our work, we propose that turbulence appears behind the shock front. Random and small-scale magnetic fields
are generated by turbulence. Mini-emitters radiate jitter photons in a bulk jet structure. Thus, the configuration of tangled magnetic
fields in a compressed slab exactly matches the physical conditions presented in our proposal. We can apply the magnetic field
configuration of \citet{laing80,laing02} to our jitter radiation process and obtain the polarization features of GRB prompt emission.

We review jitter radiation and turbulent properties in Section 2.1. The polarization feature in the case of a stochastic magnetic field
configuration is given in Section 2.2. The observed polarization quantities in a jet-in-jet scenario are presented in Section 2.3.
We briefly summarize our results in Section 3 and we present a discussion in Section 4.

\section{Polarization Processes}
GRB explosion produces relativistic shocks propagating in a bulk jet. Turbulence is thought to appear behind shocks.
Random and small-scale magnetic fields can be generated by turbulence. This kind of magnetic field is within one slab,
which is likely to be compressed by relativistic shocks. Relativistic electrons emit jitter photons in
random and small-scale magnetic fields. The linear polarization feature is determined by this specific
magnetic field topology. We sum up all the contributions of those mini-emitters in the slab and the observed
polarization degree can be finally obtained in jet-in-jet scenario. The main physical points are illustrated in Figure 2.

\subsection{Jitter Radiation and Turbulence}
We propose that jitter radiation can dominate GRB prompt emission and that
random and small-scale magnetic fields are generated by turbulence \citep{mao11}.
In this subsection, we briefly describe the major physical processes.
The radiation intensity (energy per unit frequency per unit time) of a single relativistic electron in small-scale
magnetic field was given by \cite{landau71} as
\begin{equation}
I_\omega=\frac{e^2\omega}{2\pi
c^3}\int^{\infty}_{\omega/2\gamma_\ast^2}\frac{|{\bf w}_{\omega'}|^2}{\omega'^2}{(1-\frac{\omega}{\omega'\gamma_\ast^2}+\frac{\omega^2}{2\omega'^2\gamma_\ast^4})}d\omega',
\end{equation}
where $\gamma_\ast^{-2}=(\gamma^{-2}+\omega^2_{pe}/\omega^2)$,
$\omega'=(\omega/2)(\gamma^{-2}+\theta^2+\omega^2_{pe}/\omega^2)$,
$\omega_{pe}$ is the background plasma frequency, $\theta\sim 1/\gamma$ is the angle between the electron velocity and the radiation
direction, $\gamma$ is the electron Lorentz factor, $\omega$ is the radiative frequency, and ${\bf w}_{\omega'}$ is the Fourier transform of
electron acceleration term.
In order to calculate the averaged acceleration term, a Fourier transform of the Lorentz force should be performed. The random and small-scale magnetic fields are introduced in the Lorentz force.
Following the treatment of \citep{fleishman06} and \citet{mao11}, we further obtain the
jitter radiation feature as
\begin{equation}
I_\omega=\frac{e^4}{m^2c^3\gamma^2}\int^{\infty}_{1/2\gamma_\ast^2}d(\frac{\omega'}{\omega})(\frac{\omega}{\omega'})^2{(1-\frac{\omega}{\omega'\gamma_\ast^2}
+\frac{\omega^2}{2\omega'^2\gamma_\ast^4})}
\int{dq_0d{\bf q} \delta(w'-q_0+{\bf qv})K({\bf q})\delta[q_0-q_0({\bf q})]},
\end{equation}
where the term of $K(q)$ is related to the random magnetic field. The dispersion relation $q_0=q_0({\bf q})$ is in the fluid field, $q$ and $q_0$ are the wave-number and frequency of the disturbed fluid field, respectively, $v$ is the electron velocity introduced in perturbation theory, and the radiation field can be linked
with the fluid field by the relation $\omega'=q_0-{\bf qv}$.
We adopt the dispersion relation of relativistic collisionless shocks presented by \citet{mi06} and
find that $q_0=cq[1+\sqrt{1+4\omega_{pe}^2/\gamma c^2q^2}/2]^{1/2}$. The relativistic electron frequency
is $\omega_{pe}=(4\pi e^2n/\Gamma_{sh}m_e)^{1/2}=9.8\times 10^9 \Gamma_{sh}^{-1/2}~\rm{s^{-1}}$, where we take the value of
$n=3\times 10^{10}~\rm{cm^{-3}}$ as the number density in relativistic shocks, and $\Gamma_{sh}$ is the shock Lorentz factor.
Thus, we have $\gamma c^2q^2 \gg \omega_{pe}$ for the case of GRB prompt emission.
Finally, we obtain the relation $\omega=\gamma^2vq(1-cos\theta_k)$,
where $\theta_k$ is the angle between the electron velocity and the fluid field direction. If $v\sim c$, we have a radiation
frequency range of $cq/2<\omega<\gamma^2cq$.

The stochastic magnetic field $<B(q)>$ in Equation (2) generated by
turbulent cascade can be given by
\begin{equation}
K({\bf q})\sim <B^2({\bf q})>\sim \int_{\bf q} ^{\infty} F({\bf q'})d{\bf q'},
\end{equation}
where $F({\bf q})\propto {\bf q}^{-\zeta_p}$ and $\zeta_p$ is the index of the turbulent energy spectrum. {\bf q} is within the range $q_\nu<q<q_\eta$.
$q_\nu$ is linked to the viscous dissipation and $q_\eta$ is related to the magnetic resistive
transfer. The Prandtl number ${\rm{Pr}}=10^{-5}T^4/n$ constrains
the value of $q$ by $q_\eta/q_\nu={\rm{Pr}}^{1/2}$ \citep{schekochihin07}, where $aT/m_ec^2=\Theta$, $\Theta \sim \Gamma_{sh}$
and $a$ is the Boltzmann constant. We adopt a length scale of viscous eddies
\citep{kumar09,narayan09,lazar09} for GRB prompt emission and obtain
$q_\nu=2\pi
l^{-1}_{\rm{eddy}}=2\pi(R/\Gamma_{\rm{sh}}\Gamma_t)^{-1}=6.3\times
10^{-10}(R/10^{13}\rm{cm})^{-1}(\Gamma_{\rm{sh}}/100)(\Gamma_t/10)~\rm{cm^{-1}}$,
where $\Gamma_t$ is the Lorentz factor of
turbulent eddies. The magnetic resistive scale is $q_\eta=3.9\times 10^4(n/3\times 10^{10}
\rm{cm^{-3}})^{-1/2}(T/5.6\times
10^{11}\rm{K})^2~\rm{cm^{-1}}$.
In the compressed two-dimensional case, magnetic fields can be $<B>=[\int_{q_{\nu}}^{q_\eta} {\bf q}^{-\zeta_p}d{\bf q}]^{1/2} =
\sqrt{2\pi} q_{\nu}^{(2-\zeta_p)/2}/\sqrt{\zeta_p-2}$ under the condition $q_\eta\gg q_\nu$.
Through the cascade process of a turbulent fluid, turbulent
energy dissipation has a hierarchical fluctuation structure.
A set of inertial-range scaling laws of fully developed turbulence
can be derived. From the research of \citet{she94} and
\citet{she95}, the energy spectrum index $\zeta_p$ of
turbulent fields is related to the cascade process number $p$ by
the universal relation of $\zeta_p=p/9+2[1-(2/3)^{p/3}]$. The
Kolmogorov turbulence is presented as $\zeta_p=p/3$.
This turbulent feature was found recently to be valid in the relativistic regime \citep{zrake12}.
At the fireball radius of
$R\sim 10^{13}$ cm, taking the turbulent spectrum index of $\zeta_p=3.25$ from \citet{she94}, we obtain a 
the magnetic field value of $1.3\times 10^6$ G. In the fireball scenario, a magnetic field
of $10^{14}$ G at $10^6$ cm can easily reach $10^6$ G at $10^{14}$ cm, providing powerful prompt
emission \citep{piran05}. This estimation is dependent on the exact shock location
and on the equipartition parameters. In our model, we take a magnetic field number
of $10^6$ G as a reference value. As we have shown in our model, the random magnetic field
is generated by turbulence and the magnetic field number is dependent on the turbulent
energy spectral index $\zeta_p$. 
Therefore, we solve
Equation (2) and obtain the radiation property $I_\omega\propto \omega^{-(\zeta_p-2)}$.
This radiation spectrum can be reproduced by the numerical calculations of \citet{teraki11} in high-frequency regime.
The gross radiative emission should be the contribution from all of the relativistic electrons
with a certain electron energy distribution. However, we note that the jitter spectral index
$\zeta_p-2$ of single electrons is fully determined by the fluid turbulence. Thus, the spectral index of the gross radiative emission
is not related to the electron energy distribution.
Finally, from Equation (2), we obtain
\begin{equation}
I_\omega\propto B^2.
\end{equation}
We note that the term of the magnetic field is not related to the spectral index.

\subsection{Polarization}
Magnetic field configurations is a dominant issue for the research of GRB radiation and polarization.
In this paper, we apply the topology of random magnetic fields introduced by \citet{laing80,laing02}.
A three-dimensional cube containing random and small-scale magnetic fields can be compressed into a two-dimensional slab by
relativistic shocks.
In the slab plane, the distribution of magnetic fields is still stochastic.
The magnetic field vector at one point, denoted in rectangular coordinates, is
${\bf B}=B_0(\cos\phi \sin\theta_B, \sin\phi, \cos\phi \cos\theta_B)$, where
$\theta_B$ is the angle between the slab plane and the line of sight (see Figure 1) and $\phi$ is the azimuthal angle at any point randomly distributed
in the slab plane. The position angle of the $E$-vector $\chi$ can be given as $\cos 2\chi=-(\sin^2\theta_B-\tan^2\phi)/(\sin^2\theta_B+\tan^2\phi)$.
Thus, the magnetic field acting on the radiation is $B=B_0(\cos^2\phi \sin^2\theta_B+\sin^2\phi)^{1/2}$.
Because an electron obtains same acceleration on average from different directions in a random and small-scale magnetic field, the electron trajectory of jitter radiation can be approximated as a straight line \citep{medvedev00,medvedev11}. Thus, jitter radiation is limited to the small radiation cone along the line of sight to the observer.
Since the acceleration term of jitter radiation is proportional to $B^2$, we can do a decomposition of jitter radiation in the electron radiation plane and obtain the polarization degree (see the detailed examination in the Appendix).
If the slab is orientated so that the $E$-vector of the polarization radiation has a
position angle of zero degrees, then the Stokes parameter $U=0$.
Following the magnetic field topology given by \citet{laing80,laing02}, we obtain Stokes parameters of single electron jitter radiation given by:
%\begin{equation}
$I=C\int_0^{2\pi}(\cos^2\phi \sin^2\theta_B+\sin^2\phi)d\phi$,
%\end{equation}
%\begin{equation}
$Q=I\rm{cos}2\chi=-C\int_0^{2\pi}(\cos^2\phi \sin^2\theta_B-\sin^2\phi)d\phi$,
%\end{equation}
%\begin{equation}
$U=0$,
%\end{equation}
and
%\begin{equation}
$V=0$,
%\end{equation}
where $C=C(\gamma, B_0)$.
With a certain electron energy distribution $N(\gamma)$, we obtain the following Stokes parameters of the gross jitter radiation:
\begin{equation}
I=\int^{\gamma_2}_{\gamma_1}C(\gamma, B_0)N(\gamma)d\gamma \int_0^{2\pi}(\cos^2\phi \sin^2\theta_B+\sin^2\phi)d\phi,
\end{equation}
\begin{equation}
Q=-\int^{\gamma_2}_{\gamma_1}C(\gamma, B_0)N(\gamma)d\gamma \int_0^{2\pi}(\cos^2\phi \sin^2\theta_B-\sin^2\phi)d\phi,
\end{equation}
\begin{equation}
U=0,
\end{equation}
and
\begin{equation}
V=0.
\end{equation}
Thus, the final polarization degree of the gross jitter emission is
\begin{equation}
\Pi=\frac{Q}{I}=-\frac{\int_0^{2\pi}(\cos^2\phi \sin^2\theta_B-\sin^2\phi)d\phi}{\int_0^{2\pi}(\cos^2\phi \sin^2\theta_B+\sin^2\phi)d\phi}.
\end{equation}
In our case, we emphasize that the magnetic field configuration is the only parameter that impacts the jitter
polarization. Moreover, this polarization result is only valid in the jitter radiation case where the electron deflection angle is small (see Equation (11) in Section 4). Thus, in this two-dimensional case, we only select electrons moving roughly perpendicular to the slab plane.

We present synchrotron polarization of a single electron applying the same magnetic field topology in the Appendix.
We also repeat the Stokes parameters of the gross synchrotron radiation given by \citet{laing80}:
$I=C(\gamma, B_0)\int_0^{2\pi}(\cos^2\phi \sin^2\theta_B+\sin^2\phi)^{(\nu_S+1)/2}d\phi$,
$Q=-C(\gamma, B_0)(\frac{3\nu_S+3}{3\nu_S+5})\int_0^{2\pi}(\cos^2\phi \sin^2\theta_B+\sin^2\phi)^{(\nu_S-1)/2}(\cos^2\phi \sin^2\theta_B-\sin^2\phi)d\phi$, $U=0$, and $V=0$, where $\nu_S$ is the synchrotron spectral index. We compare synchrotron polarization and jitter polarization in Figure 2.
In this paper, we neglect the effect of light propagation in a thick and highly magnetized plasma
screen \citep{macquart00}.

\subsection{Jet-in-jet Scenario}
The mini-jets emit photons in a bulk GRB jet. We simply take into
account the geometric effect of the bulk jet. The solid angle element of the jet is
$2\pi$sin$\theta$$d\theta$. Since observers always see the forward jet and the backward jet is
not observed, the detection probability is half of the result of above the estimation. As the full
opening angle is $\theta$, we suggest an upper limit of the integral to be $\theta/2$. The probability
should be normalized by the entire solid angle of $4\pi$. Thus,
the observational probability of these mini-jets in a bulk GRB jet can be calculated as
$P=\pi\int_0^{\theta/2}\sin\theta' d\theta'/4\pi=1/32\Gamma^2$ and $\Gamma \sim \theta$. The gross Lorentz factor $\Gamma$
was given by \citet{giannios10} as $\Gamma=2\Gamma_j\Gamma_t/\alpha^2$, where $\Gamma_j\sim \Gamma_{sh}$ is the Lorentz factor of
the bulk jet, $\Gamma_t$ is the Lorentz factor of relativistic turbulence in these mini-jets, and textbf{$\alpha$ is the ``off-axis''
parameter defined as $\theta_j=\alpha/\Gamma_j$. From the investigation of the mini-jets emitting angle distribution \citep{giannios10}, 
the range of $\alpha$ is given as $0<\alpha<2$. We can furthermore estimate the number of mini-jets affected by the turbulent fluid as
$n(\gamma)=4\pi R^2\Gamma_jct_{\rm{cool}}/l_s^3$, where $R$ is the fireball radius and $t_{\rm{cool}}=6\pi m_ec/\sigma_T\gamma B^2$ is
the cooling timescale of relativistic electrons. The length scale of those mini-jets is
$l_s=\gamma \Gamma_tct_{\rm{cool}}$. As the number of mini-jets $n(\gamma)$ is a function of the electron Lorentz factor $\gamma$, the
electron energy distribution $N_e(\gamma)$ is required to obtain a total number of mini-jets given by $n=\int^{\infty}_1 n(\gamma)N_e(\gamma)d\gamma/\int^{\infty}_1 N_e(\gamma)d\gamma$. Due to diffusive shock acceleration, electrons can be accelerated and a power-law
energy distribution is given. In our paper, turbulence is considered not only to generate magnetic fields but also to
accelerate electrons and the electron energy distribution has a Maxwellian component \citep{stawarz08}. If turbulent acceleration is considered
in addition to shock acceleration, the electron energy distribution has a dual-natured shape
with a Maxwellian component and a power-law component \citep{giannios09}:
\begin{equation}
 N_e(\gamma)= \left\{
\begin{array}{l l}
C\gamma^2\rm{exp}(\gamma/\Theta)/2\Theta^3, & \gamma \le \gamma_{th}, \\
C\gamma_{th}^2\rm{exp}(\gamma_{th}/\Theta)(\gamma/\gamma_{th})^{p_e}/2\Theta^3, & \gamma> \gamma_{th}, \\
\end{array} \right.
\end{equation}
where $C$ is a constant. In the case that only a fraction of electrons have a non-thermal distribution, we take the characteristic temperature
to be $\Theta=kT/m_ec^2\sim 100$. In our calculation, we adopt $\gamma_{th}=10^3$; $p_e=2.2$ is the power-law index.
Combined with the observational probability $P$ and the total number $n$ of mini-jets affected by turbulence, the final observed
polarization degree is $\Pi_{obs}=nP\Pi$.

\section{Results}
The intrinsic polarization results of GRB prompt emission without jet effects are shown in Figure 2. The jitter polarization degree is strongly
dependent on $\theta_B$, which is the angle between the line of sight and the slab plane. If the line of sight is perpendicular to the
slab plane, we successfully obtain jitter photons, but jitter radiation in the electron radiative plane is symmetric and the polarization degree is zero. Except in this extreme case, the
degree of polarization can be measured with different values of $\theta_B$. With the same magnetic field configuration, we can compare
jitter polarization and synchrotron polarization. Both jitter polarization and synchrotron polarization are related to $\theta_B$. On the
other hand, as we discussed in Section 2.2, jitter polarization is not related to the radiation spectrum and it can reach a maximum value of
100 percent. Synchrotron polarization (see details in the Appendix) is related to the radiation spectral index and its maximum value is 
$(\nu_S+3)/(\nu_S+5)$. The cases of $\nu_s=2.0,1.0$, and 0.6 are shown in Figure 2 as examples.

We further provide examples of the observed jitter polarization degree in our jet-in-jet scenario shown in Figures 3 and 4.
Different jet ``off-axis'' parameters, 1.55, 1.3 and 1.0, are used to calculate the observed polarization degree shown in Figure 3. Here, we take $\Gamma_{sh}=100$, $\Gamma_t=10$, and $R=10^{13}$ cm.
We show that the polarization degree is significantly sensitive to the jet ``off-axis'' parameter.
A stronger jet ``off-axis'' effect yields a stronger observed polarization degree.
In Figure 4, we present examples of jitter polarization results affected by the shock Lorentz factor $\Gamma_{sh}$.
We fix the ``off-axis'' parameter as 1.3, $\Gamma_t=10$, and $R=10^{13}$ cm.
Shock Lorentz factor values of 85, 100, and 120 are adopted.
We see that a low polarization degree can be induced by a large shock Lorentz factor. 
This fact suggest that a large shock Lorentz factor provides powerful radiation
along the jet axis while a powerful polarization can be measured in the strong off-axis case, which
has a relatively small Lorentz factor and weak radiation.

It is also useful to review the GRB polarization feature under the mechanism of synchrotron radiation in the jet structure.
\citet{gruzinov99b} and \citet{sari99} proposed the possibility of high-degree GRB polarization. \citet{granot03}
comprehensively discussed different GRB jet polarization cases. Here, for simplicity, we apply the model of \citet{gruzinov99b}
to illustrate the configuration differences between the tangled magnetic field and the ordered magnetic field.
The polarization degree derived from the magnetic field parallel/perpendicular to the jet direction was given as
$\Pi=\Pi_0\sin^2\alpha_B(B^2_\parallel-0.5B^2_\perp)/[B^2_\parallel\sin^2\alpha_B+0.5B^2_\perp(1+\cos^2\alpha_B)]$, where $B_\parallel$ is
the magnetic field parallel to the shock propagation direction, $B_\perp$ is the magnetic field perpendicular to the shock propagation,
and $\alpha_B=\pi/2-\theta_B$ is the angle between the line of sight from observer and the direction of the shock propagation \citep{gruzinov99b}.
In our model, we
have fully obtained the polarization feature of GRB prompt emission in the case of $B_\parallel\ll B_\perp$ \citep{laing80,laing02} and this condition is necessary for jitter radiation. If
$B_\parallel\gg B_\perp$, the polarization degree is $\Pi \sim \Pi_0$, which is the result obtained from an ordered magnetic field
aligned with the jet direction. This result reproduces exactly the polarization degree of synchrotron radiation as $\Pi_0=(3\nu_S+3)/(3\nu_S+5)$;
the electron energy distribution has a power-law distribution with a power-law index $p_e$ and $\nu_S=(p_e-1)/2$.
Considering a certain jet structure and line of sight effects, a high degree of GRB prompt polarization can be
obtained \citep{granot03,lazzati09}.
In our paper, the two-dimensional magnetic slab and the jet-in-jet structure provide the possibility of asymmetric radiation in the electron radiative plane. Thus, the GRB prompt emission can be highly polarized as well.

\section{Discussion and Conclusions}
Relativistic electrons accelerated by diffusive shock acceleration with a power-law energy distribution can emit high-energy
photons in an ordered magnetic field. Therefore, it is possible to explain the high-degree polarization detected in GRB prompt
emission by synchrotron radiation.  Alternatively, we present in this paper that turbulence can accelerate electrons and generate random and
small-scale magnetic fields. The relativistic electrons emit X-ray/$\gamma$-ray jitter photons in the jet-in-jet structure. Some important
physical components, such as turbulence behind the shock front, random and small-scale magnetic fields, the electron energy distribution,
and mini-jets, are self-consistently organized in the process of jitter radiation. We obtain a high-degree
polarization of GRB prompt emission through jitter radiation considering a
two-dimensional compressed magnetic field configuration.
The intrinsic polarization degree is a function of the angle
between the line of sight and the slab plane, and large polarization values can be achieved in the case of small angles.
Moreover, the final observed polarization degree in the jet-in-jet scenario is affected by the ``off-axis'' parameter and the shock
Lorentz factor.

In this paper, the electron trajectory of jitter radiation in random and small-scale magnetic fields is assumed to be a straight line. In other words, the condition of $\theta_d\ll 1/\gamma$ should be satisfied by jitter radiation \citep{medvedev11}. Here, $\theta_d=eB\lambda_B/\gamma m_ec^2$ is the deflection angle of electrons in a magnetic field, and $\lambda_B\sim c/\omega_{pe}$ is the length scale of the magnetic field. We obtain
\begin{equation}
\theta_d=(\frac{B}{1.2\times10^6~\rm{G}})(\frac{n}{1.5\times10^{19}~\rm{cm^{-3}}})^{-1/2}(\frac{\Gamma_{sh}}{100})^{1/2}\gamma^{-1}.
\end{equation}
This condition is strongly dependent on the magnetic field and the electron number density. We constrain the condition of jitter radiation in an example below. The electron number density is given by $n=N/\Delta\Omega R^2\Delta R$. $N=1.0\times 10^{52}\sim 6\times 10^{54}$ is the total number of electrons in the radiative region \citep{kumar09}. We take the fireball radius to be $R=10^{13}~\rm{cm}$ and fireball shell to be $\Delta R=10^{10}~\rm{cm}$. If the jet opening angle is about 5 degree, we obtain a solid angle $\Delta\Omega= 3.3\times 10^{-4}$ and the estimated value of number density is $n=3.0\times 10^{20}~\rm{cm^{-3}}\sim 1.8\times 10^{22}~\rm{cm^{-3}}$. Thus, $\theta_d\ll 1/\gamma$ and the electron trajectory can be treated as a straight line.
In order to satisfy this jitter condition in the two-dimensional case, we select electrons moving roughly perpendicular to the slab plane so that the electron deflection angle is small. The general case of electrons moving in a random walk in a stochastic magnetic field with a large deflection angle was discussed by \citet{fleishman06} and numerical simulations may be required to solve this complicated issue.

It is expected that the so-called depolarization due to the Faraday rotation of the polarization screen \citep{burn66} can be a powerful tool to
further constrain the source structure and magnetic field topology. In general, stochastic Faraday rotation creates a polarization
degree of $\Pi\propto \rm{exp}(-\lambda^4)$, where $\lambda$ is wavelength \citep{melrose98}. \citet{tribble91} discussed a power-law structure
function for a Faraday rotation measurement. The polarization degree was given as $\Pi\propto \lambda^{-4/m}$, where $m$ is related to the
turbulent cascade. Even if we consider a polarization fluctuation proportional to $\lambda$, the
final polarization degree still dramatically decreases as $\Pi\propto \lambda^{-2.2}$ if we adopt a typical value of $m=\zeta_p-2$ given
in our model. Therefore, GRB prompt polarization measured in the soft band is about 20\% of that measured in the hard band. This
prediction can be examined by future observations if GRB prompt polarization measurements can be performed simultaneously in different
energy bands. This simple analysis of the depolarization effect also indicates that GRB prompt polarization in optical bands is nearly impossible detect even if GRB is highly polarized in high-energy bands. On the other hand, the polarization of GRB optical
afterglows was detected. The radiation case of external shocks of GRBs sweeping into the surrounding
interstellar medium is beyond the scope of this paper. Here, we make some simple speculations. As optical afterglow is onset at
early times after the GRB is triggered, and the jet is strongly anisotropic with a narrow beaming angle. Thus, the optical polarization degree can reach
values of 10\% \citep{uehara12} and even higher polarization degrees were expected \citep{steele09}. In late times after a GRB is triggered,
the beaming angle of the jet is wide and the jet anisotropy is not prominent. Thus, the polarization of the optical afterglow is weak\footnote{
The only abnormal case is GRB 020405. About 10\% of the polarization was still measured 1.3 days after the GRB was triggered \citep{bersier03}.}
\citep{covino99,hjorth99,wijers99}.  Moreover, high-energy photons from GRB prompt emission
can pass through a dense medium without being strongly absorbed. Thus, the depolarization effect is weak and a high
polarization degree can be detected, while UV/optical photons of GRB afterglows can be strongly absorbed by the surrounding dense material.
Therefore, even though the initial polarization degree of GRB optical afterglows is very high, the external depolarization effect is strong as
the diffraction in a Faraday prism is taken into account \citep{sazonov69,macquart00}. Thus, due to the strong depolarization effect,
the detected polarization degree of GRB optical afterglows is low. Further quantitative computations can be executed since the magnetic field
configuration adopted in this paper is still valid for the depolarization cases mentioned above \citep{rossi04}.

Rapid time variations of the polarization angle and polarization degree have been measured either in GRB prompt emission cases
(\citet{gotz09} for GRB 041219A and \citet{yonetoku11} for GRB 100826A) or in GRB afterglow cases (\citet{rol00} for GRB 990712,
\citet{greiner03} for GRB 030329, and \citet{wier12} for GRB 091018). The observed timescale of polarization variation in GRB prompt
emission is about 50-100 s \citep{yonetoku11}, which corresponds to an intrinsic timescale of 0.5-1 s.
Here, we list three possibilities to explain this rapid variation. First, GRB helical jets with helical magnetic fields and helical rotation generated by black holes \citep{mizuno13} should be considered. With a helical jet, the slab
may rotate and the magnetic field configuration may change quickly. Then, rapid polarization variation occurs. Second,
\citet{lazzati09} proposed that faint pulses of GRB prompt emission are a main contributor to the polarization. If this is true, we can obtain
the rapid time variation of the polarization since the pulses shown in GRB prompt lightcurves show rapid variation;
Third, the turbulent feature in the slab should be revisited. We estimate the variability of GRB prompt polarization due to the variability of
stochastic eddies as $\delta t\sim l_{eddy}/c_s$.
Given $l_{eddy}\sim R/\Gamma_j\Gamma_t$ and $c_s\sim c/\sqrt{3}$, $\delta t$ is about 1.7 s, which can be comparable to the observational
timescale. Moreover, small-scale fluid turbulence and magnetic reconnection was suggested by
\citet{matthews90}. Recently, \citet{zhang11} and \citet{mckinney12} illustrated the possibility of magnetic reconnection for GRB energy dissipation. We suggest that small-scale magnetic reconnection in the compressed slab is likely to affect 
the rapid polarization variation. More observations (TSUBAME, NuSTAR, Astro-H) and quantitative explanations in detail are expected in the future.

\acknowledgments
We thank the referee for his/her helpful suggestions and comments. We are grateful to Teraki, Y., Toma, K., Nagataki, S., Ito, H., Ono, M., and Lee, S.-H. for their useful discussions. The numerical computation in this work was carried out at the Yukawa Institute
Computer Facility. This work is supported by Grants-in-Aid for Foreign JSPS Fellow (Number 24.02022). We acknowledge support from the
Ministry of Education, Culture, Sports, Science and Technology (No. 23105709), the Japan Society for the Promotion of Science (No. 19104006 and No. 23340069), and the Global COE Program, The Next Generation of Physics, Spun from University and Emergence, from MEXT of Japan.
J. Wang is supported by the National Basic Research Program of
China (973 Program 2009CB824800), the National Natural Science Foundation of China 11133006, 11163006, 11173054, and the Policy Research Program of the Chinese Academy of Sciences (KJCX2-YW-T24).

\appendix
\section{Polarization of Single Electrons in A Random Magnetic Field}
In this Appendix, we first introduce the polarization feature of single relativistic electrons in a magnetic field assuming jitter condition.
We repeat the radiation property by \citet{landau71} as
\begin{equation}
E_\omega=\frac{1}{2\pi}\int {Ee^{i\omega t}dt},
\end{equation}
where the retarded time is $t'=t-R(t')/c\cong t-R_0/c+{\bf n}\cdot {\bf r} (t')/c\cong t-R_0/c+{\bf n}\cdot {\bf v}t'/c$; ${\bf n}$ is the radiation direction. If we take
$dt=dt'(1-{\bf n}\cdot{\bf v}/c)$, we obtain
\begin{equation}
E_\omega=\frac{e}{c^2}\frac{e^{ikR_0}}{R_0(1-{\bf n}\cdot {\bf v}/c)^2}\int {\bf n}\times\{({\bf n}-{\bf v}/c)\times {\bf w}(t')\}e^{i\omega t'(1-{\bf n}\cdot{\bf v}/c)}dt'.
\end{equation}
If we adopt the notation $\omega'=\omega(1-{\bf n}\cdot {\bf v}/c)$, we obtain
\begin{equation}
E_\omega=\frac{e}{c^2}\frac{e^{ikR_0}}{R_0}(\frac{\omega}{\omega'})^2 {\bf n}\times\{({\bf n}-{\bf v}/c)\times {\bf w}_{\omega'}\}.
\end{equation}
The acceleration term is ${\bf w_{\omega'}}=e{\bf v}\times {\bf B}/\gamma m_ec$. The electron trajectory is necessary to obtain the acceleration term and the calculation in Equation (A3). If the electron trajectory can be treated as a straight line in a magnetic field (see Equation (11) of jitter radiation condition), we have a compressed magnetic slab plane with an angle of $\theta_B$ to the radiative $X-Y$ plane, and the electron velocity is perpendicular to the slab plane since the electron deflection angle of the jitter radiation is smaller than $1/\gamma$. The jitter radiation direction follows the direction of the electron. Thus, we obtain the $X$-axis component as
\begin{equation}
E_x=\frac{e^2B_y}{2\gamma^3m_ec^2}(\frac{\omega}{\omega'})^2\frac{e^{ikR_0}}{R_0}
\end{equation}
and the $Y$-axis component as
\begin{equation}
E_y=-\frac{e^2B_x}{2\gamma^3m_ec^2}(\frac{\omega}{\omega'})^2\frac{e^{ikR_0}}{R_0}.
\end{equation}
The radiation intensity is $I=c|E|^2R_0^2/2\pi$. The Stokes parameters can be defined as
\begin{equation}
I=<E_xE^{\ast}_x>+<E_yE_y^{\ast}>=\frac{e^4}{8\pi\gamma^6m_e^2c^3}(\frac{\omega}{\omega'})^4(B^2_x+B_y^2)
\end{equation}
and
\begin{equation}
Q=<E_xE^{\ast}_x>-<E_yE_y^{\ast}>=\frac{e^4}{8\pi\gamma^6m_e^2c^3}(\frac{\omega}{\omega'})^4(B^2_y-B_x^2).
\end{equation}
From Section 2.2, we know $B_x=B_0\rm{cos}\phi\rm{sin}\theta_B$ and $B_y=B_0\rm{sin}\phi$. Finally, we obtain the jitter polarization $\Pi=Q/I$, which is only related to the configuration of the magnetic field; Equation (9) is verified.

In this Appendix, we also review the synchrotron polarization of a single electron. 
We keep the same random magnetic field topology used in
the jitter polarization case to calculate the synchrotron polarization and we can repeat
the calculation of \citet{laing80}.
The radiation intensity of the synchrotron mechanism is
\begin{equation}
I=\frac{2\pi\sqrt{3}e^2\nu_L}{c}[\frac{\nu}{\nu_c}\int^{\infty}_{\nu/\nu_c}K_{5/3}(t)dt],
\end{equation}
where $\nu_c=(3/2)\gamma^2\nu_L$, $\nu_L=eB/(2\pi m_ec)$ is the Larmor frequency ,and $K$ is a modified Bessel function. The synchrotron polarization of a single electron is
\begin{equation}
\Pi=\frac{K_{2/3}(\nu/\nu_c)}{\int_{\nu/\nu_c}^{\infty}K_{5/3}(t)dt},
\end{equation}
related to the magnetic field, electron Lorentz factor $\gamma$ and
radiation frequency $\nu$ \citep{rybicki79}. 

In order to compare synchrotron polarization with jitter polarization, we apply the same magnetic field topology described in Section 2.2 to synchrotron polarization; our results are shown in Figure 5. In the calculation, we fix the magnetic field value at
$1.3\times 10^6~\rm{G}$. For a given $\theta_B$, the degree of synchrotron polarization increases, if the radiation
is toward a higher frequency $\nu$ and/or the electron Lorentz factor $\gamma$ tends toward smaller values. This property is 
typical of synchrotron polarization of single electrons. In Figure 5, we plot the jitter polarization as
well, which is independent of the radiation frequency $\nu$ and the electron Lorentz factor $\gamma$. Even though synchrotron polarization degrees can be coincidentally the same as the jitter polarization
degree by adjusting the parameters of $\nu$ and $\gamma$, we emphasize that jitter polarization and synchrotron
polarization are physically different because jitter radiation and synchrotron radiation are two different radiation
mechanisms.

If the synchrotron power-law spectrum with a power-law index $\nu_S$ is given and an electron energy distribution is assumed to be a power-law with an index of $2\nu_S+1$, we obtain the Stokes parameters of the gross electrons:
\begin{equation}
I=C\int_0^{2\pi}(\cos^2\phi \sin^2\theta_B+\sin^2\phi)^{(\nu_s+1)/2}d\phi,
\end{equation}
\begin{equation}
Q=-C\frac{(3\nu_S+3)}{(3\nu_S+5)}\int_0^{2\pi}(\cos^2\phi \sin^2\theta_B+\sin^2\phi)^{(\nu_s-1)/2}(\cos^2\phi \sin^2\theta_B-\sin^2\phi)d\phi,
\end{equation}
\begin{equation}
U=0,
\end{equation}
and 
\begin{equation}
V=0,
\end{equation}
where $C$ is constant. Thus, the final polarization degree of the gross synchrotron emission is
\begin{equation}
\Pi=\frac{Q}{I}=-\frac{(3\nu+3)}{(3\nu_s+5)}\frac{\int_0^{2\pi}(\cos^2\phi \sin^2\theta_B+\sin^2\phi)^{(\nu_s-1)/2}(\cos^2\phi \sin^2\theta_B-\sin^2\phi)d\phi}{\int_0^{2\pi}(\cos^2\phi \sin^2\theta_B+\sin^2\phi)^{(\nu_s+1)/2}d\phi}.
\end{equation}
These are the results given by \citet{laing80} and we also plot them in Figure 2.

\clearpage

\begin{figure}
\includegraphics[angle=0,scale=1.]{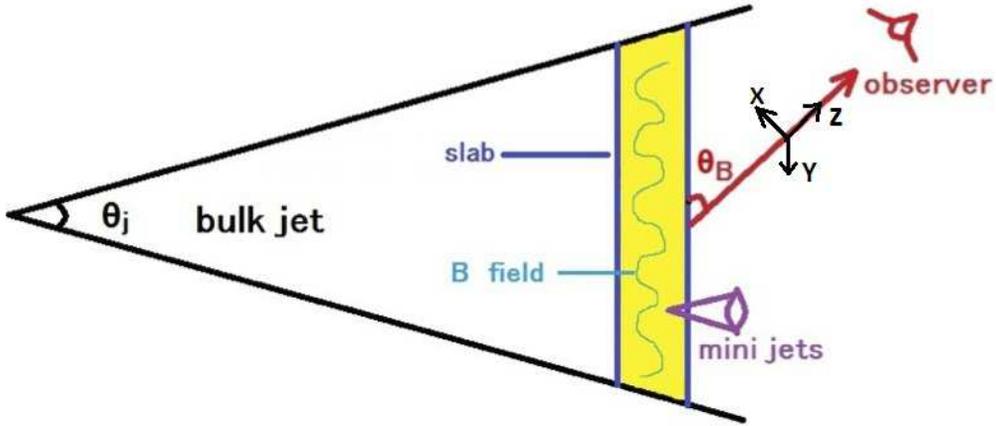}
\caption{GRB prompt emission and polarization within a compressed slab. Shock propagation is in the bulk jet structure and turbulence
occurs behind the shock front. Random and small-scale magnetic field are generated by turbulence in a three-dimensional cube. This cube can be compressed to be a two-dimensional slab.
GRB prompt emission is the total emission from those mini-jets. The GRB prompt polarization feature is dependent on the magnetic
field configuration. $B_x=B_0\rm{cos}\phi\rm{sin}\theta_B$ and $B_y=B_0\rm{sin}\phi$ are two components in the slab plane, where $\phi$ is the azimuthal angle at any point randomly distributed
in the slab plane \citep{laing80}. The angle between the line of sight from the observer and the slab plane is labeled as $\theta_B$.
\label{fig1}}
\end{figure}

\begin{figure}
\includegraphics[angle=270,scale=0.65]{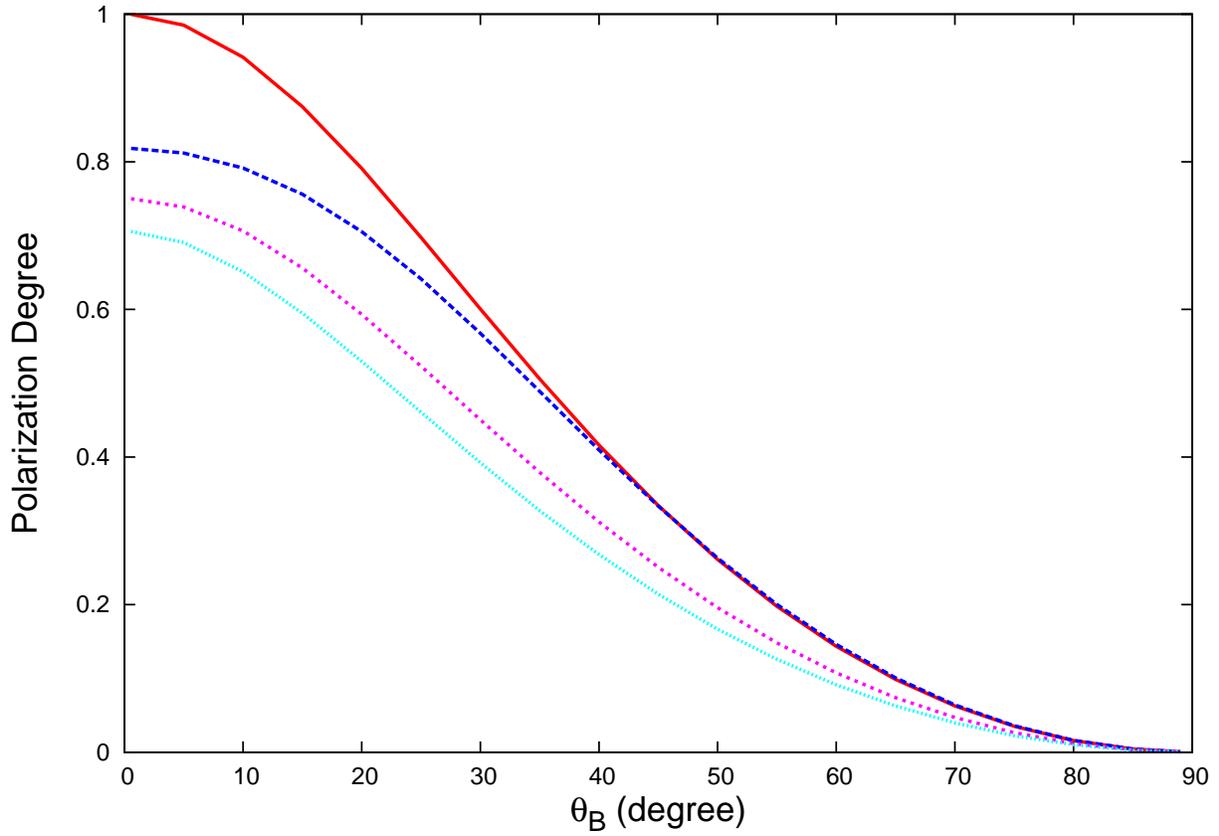}
\caption{Intrinsic polarization degree as a function of $\theta_B$. The solid line (red) denotes the polarization
result of jitter radiation. The long-dashed line (blue), the short-dashed line (pink), and the dotted line (cyan) denote the
polarization results calculated from synchrotron radiation with a spectral index of 2.0, 1.0, and  0.6, respectively.
\label{fig1}}
\end{figure}

\begin{figure}
\includegraphics[angle=270,scale=0.65]{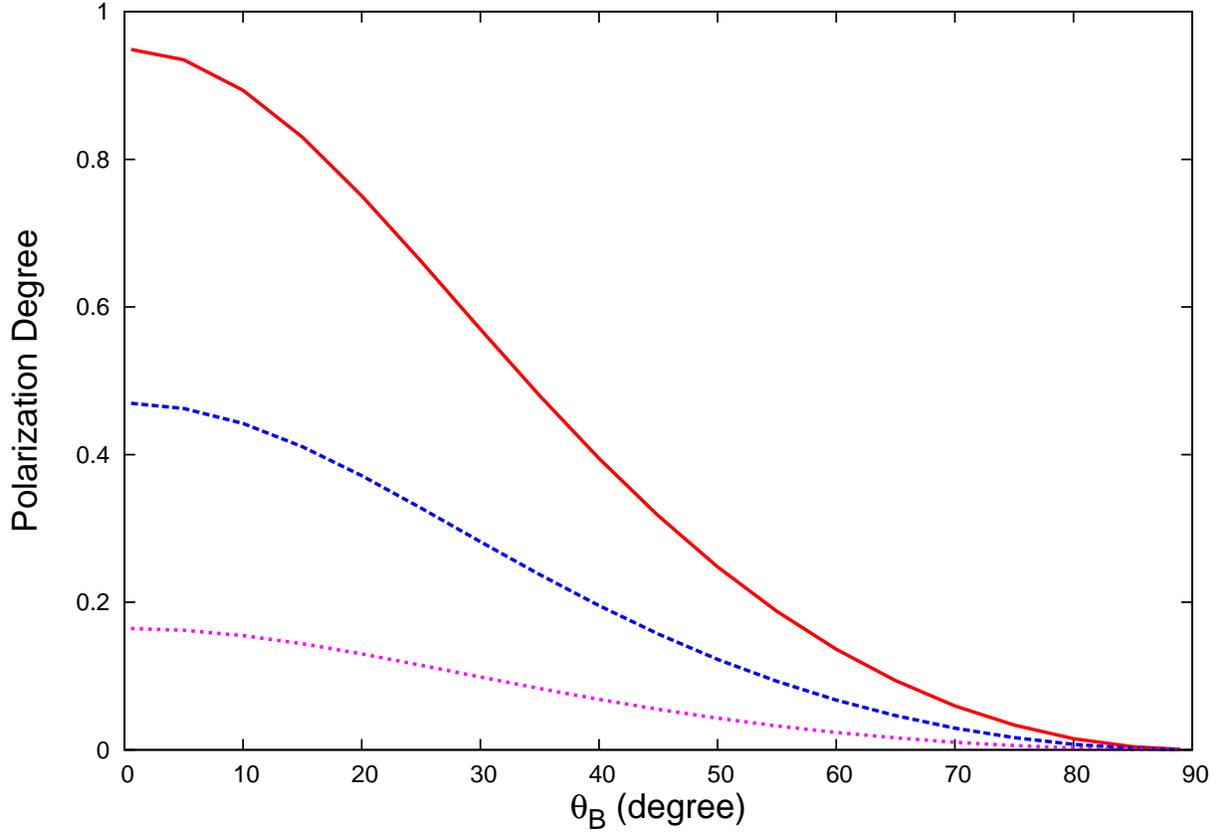}
\caption{Observational jet ``off-axis'' effect on GRB prompt polarization. The solid line (red), the long-dashed line (blue),
and the short-dashed line (pink) denote the jitter polarization results given by ``off-axis'' parameter values of 1.55, 1.3,
and 1.0, respectively. The shock Lorentz factor $\Gamma_{sh}$ is fixed at 100.
%which corresponds to $pp=14$
\label{fig1}}
\end{figure}

\begin{figure}
\includegraphics[angle=270,scale=0.65]{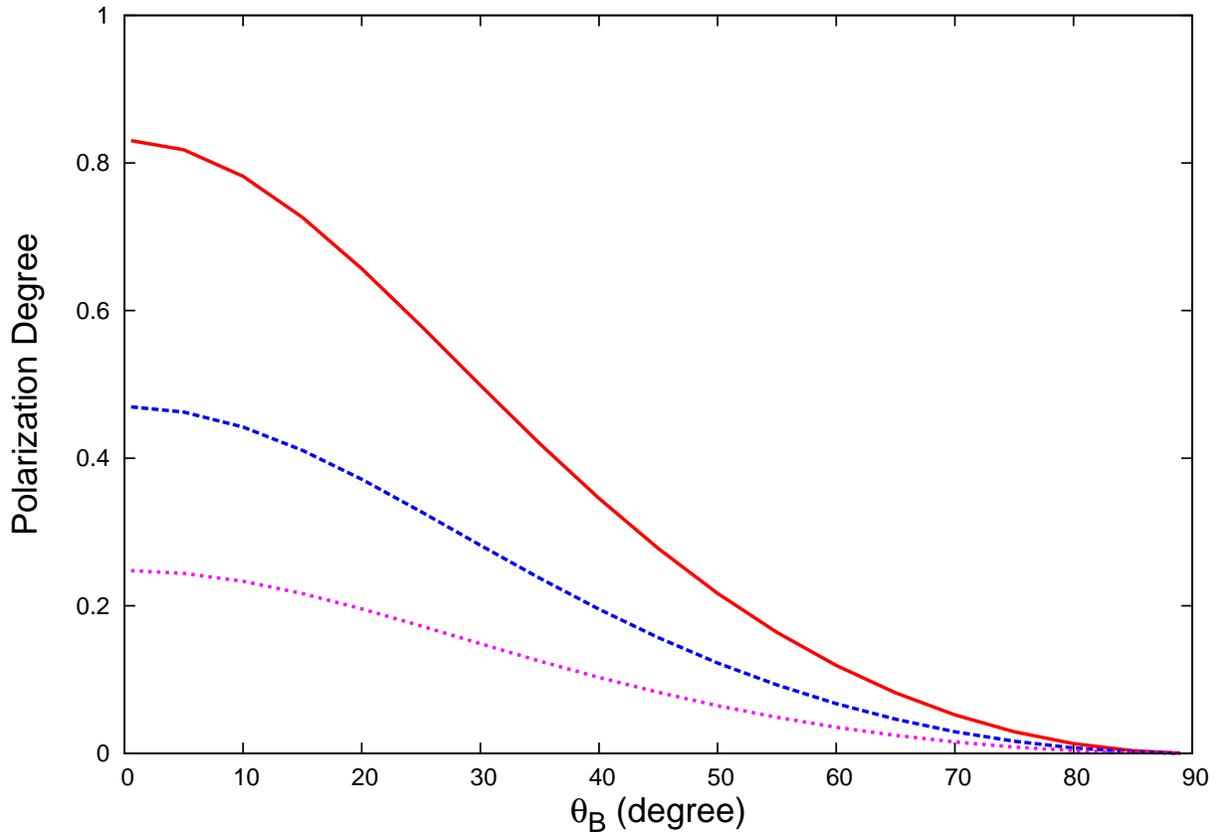}
\caption{Shock Lorentz factor effect on GRB prompt polarization. The solid line (red), the long-dashed line (blue), and
the short-dashed line denote the jitter polarization results given by shock Lorentz factors $\Gamma_{sh}$ of 85, 100, and 120,
respectively. The ``off-axis'' parameter is taken to be 1.3.
%which corresponds to $pp=14$
\label{fig1}}
\end{figure}

\begin{figure}
\includegraphics[angle=270,scale=0.5]{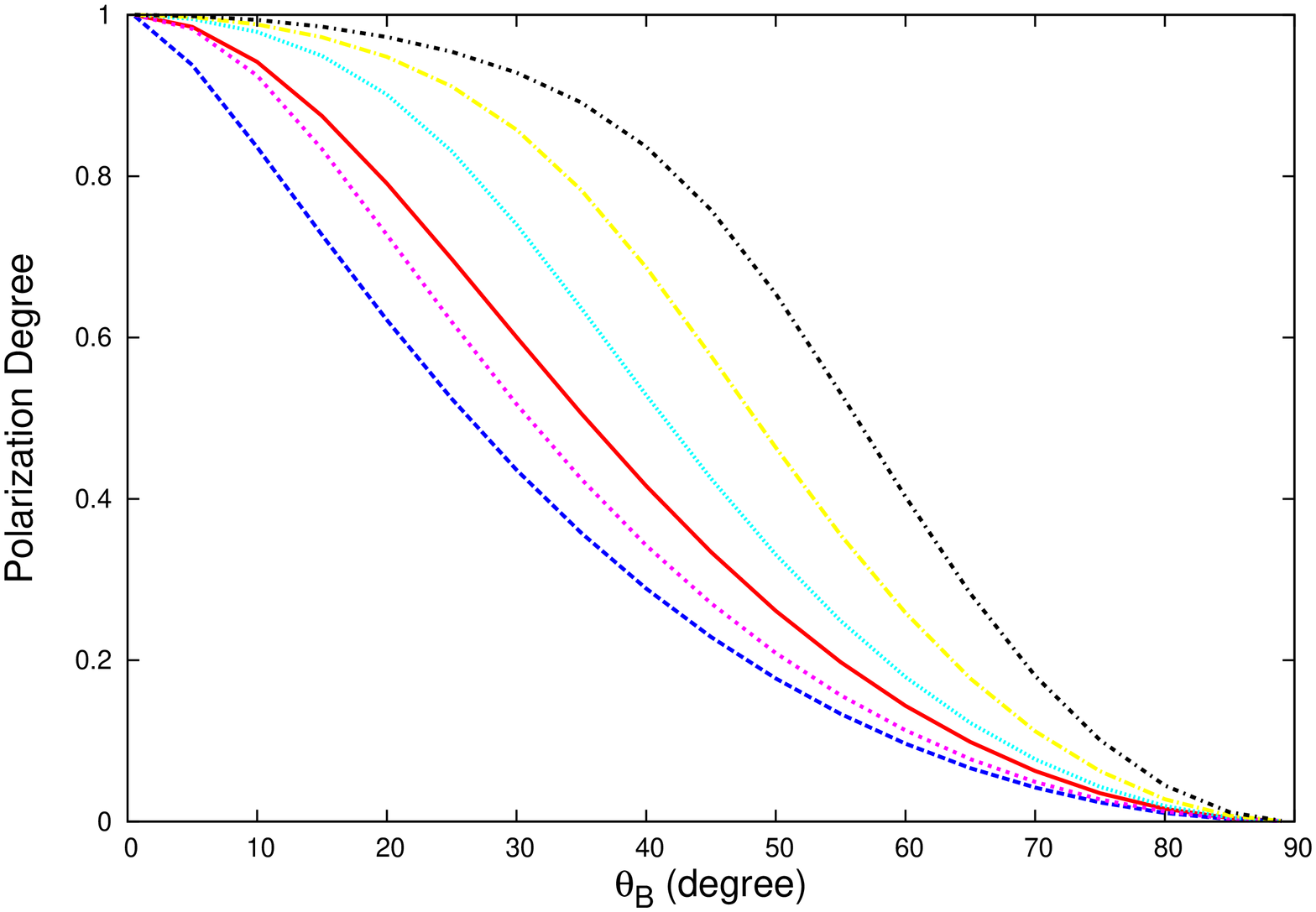}
\includegraphics[angle=270,scale=0.5]{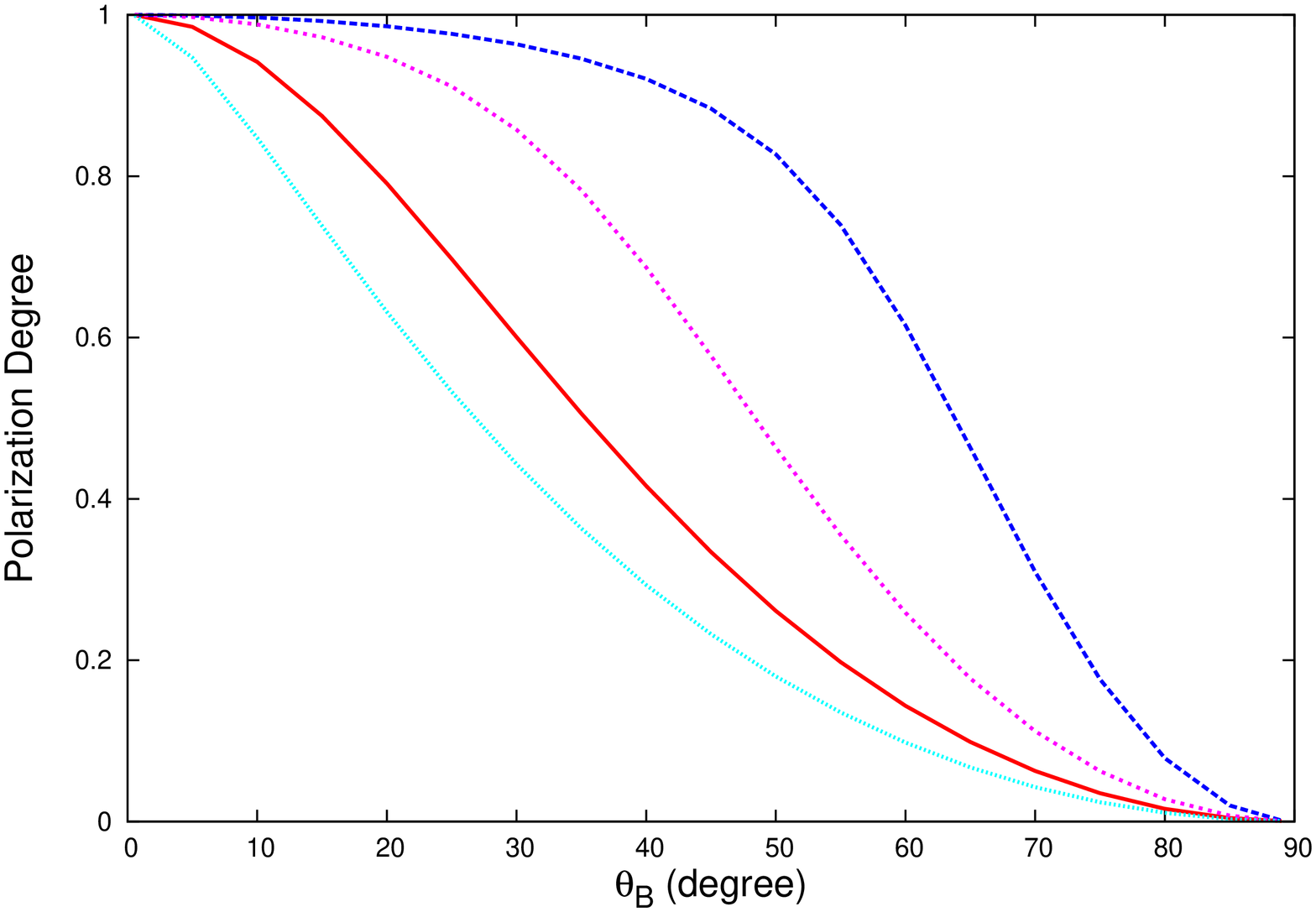}
\caption{Synchrotron polarization vs jitter polarization for a single electron. Top panel: synchrotron
polarization degree results are calculated as a function of $\theta_B$ with radiation frequencies of $\nu=1~\rm{keV}$,
$\nu=50~\rm{keV}$, $\nu=250~\rm{keV}$, $\nu=500~\rm{keV}$, and $\nu=1~\rm{MeV}$ denoted by the long-dashed
line (blue), the short-dashed line (pink), the dotted line (cyan), the long-dash-dotted line (yellow),
and the short-dash-dotted line (black), respectively. The electron Lorentz factor is fixed at $\gamma=10^3$.
Bottom panel: synchrotron polarization degree results calculated as a function of $\theta_B$ with
electron Lorentz factors $\gamma=500$, $\gamma=10^3$, and $\gamma=10^4$ denoted by the long-dashed line (blue), the short-dashed line (pink),
and the dotted line (cyan), respectively. The radiation frequency is fixed at $\nu=500~\rm{keV}$.
The jitter polarization denoted by the solid line (red) is shown in both panels as well.
\label{fig1}}
\end{figure}

\clearpage

\end{document}